\documentclass{kapproc} 

\usepackage{procps} 

\usepackage[dvips]{graphicx}

\upperandlowercase

\kluwerbib 

\let\lcitebracket(
\let\rcitebracket)

\newcommand{\gr}{\kern 2pt\hbox{}^\circ{\kern -2pt K}} 
\newcommand{\brr}{\begin{array}}
\newcommand{\err}{\end{array}}
\newcommand{\ltsima}{$\; \buildrel < \over \sim \;$}
\newcommand{\simlt}{\lower.5ex\hbox{\ltsima}}
\newcommand{\gtsima}{$\; \buildrel > \over \sim \;$}
\newcommand{\simgt}{\lower.5ex\hbox{\gtsima}}

\begin{document}

\articletitle[Comparison between optical and X-ray cluster detection
methods] 
{Optically and X-ray selected Clusters \\
of galaxies in the XMM/2df/SDSS survey}

\author{S. Basilakos \altaffilmark{1}, 
M. Plionis \altaffilmark{1,2}, 
S. Georgakakis \altaffilmark{1}, I. Georgantopoulos \altaffilmark{1}\\
T. Gaga \altaffilmark{1,3}, V. Kolokotronis \altaffilmark{1}, 
G. C. Stewart \altaffilmark{4} }

\affil{\altaffilmark{1}Institute of Astronomy \& Astrophysics, National 
Observatory of Athens, 
I.Metaxa \& B.Pavlou, Palaia Penteli, 152 36, Athens, Greece.}

\affil{\altaffilmark{2}Instituto Nacional de Astrofisica, Optica 
y Electr\'onica (INAOE) Apartado Postal 51 y 216, 72000, Puebla, Pue., Mexico.}

\affil{\altaffilmark{3}Physics Department, Univ. of Athens, Panepistimioupolis,
Zografou, Athens, Greece.}

\affil{\altaffilmark{4}Department of Physics and Astronomy, University of Leicester, UK, LE1 7RH.}

\begin{abstract}
In this work we 
present combined optical and X-ray  
cluster detection methods in an area near the 
North Galactic Pole area, previously covered
by the SDSS and 2dF optical surveys. The same area has been covered
by shallow ($\sim 1.8$ deg$^{2}$) XMM-{\em Newton} observations.
The optical cluster detection procedure is based on 
merging two independent selection methods - a smoothing+percolation    
technique, and a Matched Filter 
Algorithm. The X-ray cluster detection is based on a wavelet-based algorithm,
incorporated in the SAS v.5.2 package. The final optical sample 
counts 9 candidate clusters with richness of more than 20 galaxies, 
corresponding roughly to APM richness class. Three, of our optically 
detected clusters are also detected in our X-ray survey.
\end{abstract}


\section{Introduction}

The cosmological significance of galaxy clusters 
has initiated a number of studies aiming to compile
unbiased cluster samples to high redshifts, utilizing multiwavelength data
(e.g. optical, X-ray, radio). 
From the optical point of view there are several available 
samples in the literature
\cite{aco89, da94, ol99, goto02}
which are playing a key role in astronomical research. 
Optical surveys suffer from 
projection effects \cite{fre90} and thus, cluster detection in X-rays 
is a better approach, owing to the fact that the diffuse
Intra-Cluster Medium (ICM) emits strongly in X-rays.
The first such survey, was based on the Extended Einstein Medium 
Sensitivity Survey, containing 99 clusters  
\cite{st91}. Recently, the {\it ROSAT} satellite allowed
a leap forward in the X-ray cluster astronomy, producing large samples
of both nearby and distant clusters \cite{eb00, sch97}.

However, even with the improved sensitivity of the XMM-{\it Newton}, optical
surveys remain significantly more efficient and less expensive in 
telescope time for compiling cluster samples, albeit with some 
incompleteness and spurious detections.
The aim of this work is 
to make a comparison of optical and X-ray cluster
identification methods in order to quantify the selection biases
introduced by these different techniques and to estimate the possible
fraction of spurious optically selected clusters due to projection effects.

\section{Observations}
In this paper we utilize the SDSS Early Data Release (EDR), covering an
area of $\sim 400$deg$^2$ in the sky
\cite{stou02}. Goto et al. (2002) applied an objective cluster 
finding algorithm to the
SDSS EDR and produced a list of 4638 galaxy clusters,
with estimated photometric redshifts.
Furthermore, we analyzed 9 {\it XMM-Newton} fields with nominal exposure time
between 2 and 10 ksec, covering an area of $\rm 1.8 \;\;deg^2$. 
However, one of the fields, suffering from significantly elevated and 
flaring particle background, was excluded from the 
X-ray analysis. 
 
\subsection{The optical finding algorithms}
The first cluster detection algorithm is based on smoothing the 
discrete distribution using
a Gaussian smoothing kernel. We select all grid-cells with overdensities 
above a chosen critical threshold  ($\delta \ge 1$)
and then we use a friends-of-friends algorithm to form groups of 
connected cells, which we consider as
our candidate clusters.
Note that the grid cell size is such that at $z=0.4$ it
corresponds to 100 $h^{-1}$kpc ($\sim 19''$).
The second optical cluster detection method is the 
matched filter algorithm (hereafter MFA) described by 
Postman et al. (1996).

Finally, we construct our final cluster catalogue, in the 
$\sim 1.8$ deg$^{2}$ area covered by our XMM survey,
by adopting the conservative
approach of considering as cluster candidates those identified by
both independent selection methods (described before).
This sample, contains 9 clusters with SDSS richness of more than 20 galaxies, 
corresponding roughly to APM type clusters. 
Comparing our final list of 9 clusters with the 
Goto et al. (2002) clusters, we find 5 in common.

\subsection{X-ray Cluster Detection}
In order to detect candidate clusters in our XMM fields
we use the soft 0.3-2 keV band since it maximizes the
signal to noise ratio, especially in the case of relatively
low temperature galaxy clusters. In particular, we 
utilize the {\sl EWAVELET} detection algorithm of the {\it XMM-Newton}
SAS v.5.2 analysis software package, which 
detects sources on the wavelet transformed images.
We have detected 7 candidate clusters on the MOS mosaic, while
5 extended sources were detected on the PN images, out of which 3
overlap with the MOS candidates. After 
excluding obvious double point sources and 
MOS-edge effects
we are left with 4 X-ray candidate clusters. 
The faintest extended source has a flux of 
$\rm \sim 2 \times 10^{-14}erg ~cm^{-2}~s^{-1}$.

\begin{figure}[t]
\includegraphics[width=10.5truecm]{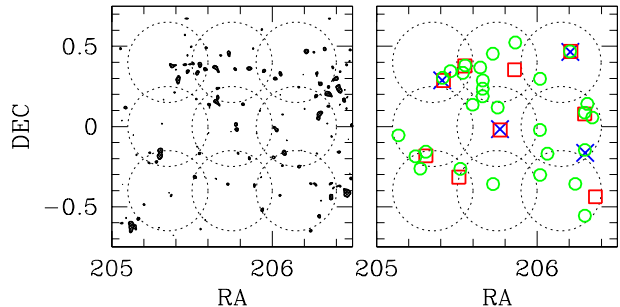}
\caption{
The smooth SDSS density field (left panel) in equatorial coordinates. 
The candidate cluster positions (right panel) within our shallow
XMM survey. Open squares are the optically selected clusters 
using our technique, while open circles are the Goto et al. (2002) 
clusters. Crosses are our X-ray selected clusters. 
The large dotted circles represents the XMM 15 arcmin radius fields of
view of our shallow XMM survey.}

\label{fig:basprof2}
\end{figure}

\section{Results}
In Fig. 1 (left panel) we plot the smoothed SDSS galaxy distribution, on 
the equatorial plane with contours 
delineating the $\delta_{\rm cr} =1$ level.
While in Fig. 1 (right panel), we plot the positions
of (a) our detected cluster candidates, 
(b) the Goto et al. (2002) clusters and (c) the
X-ray detected XMM clusters. 
All four X-ray detections coincide with optical 
cluster candidates from different
methods, with the largest coincidence rate (3 out of 9) 
being with our methods. 
The most distant cluster, $z=0.67$ \cite{couch91} in our 
optical sample, located 
at $RA =$13h 43min 4.84sec and
$DEC=$ 00$^{o}$ 00$^{'}$ 56.26$^{''}$, found also in 
X-rays, is missed by Goto et al. (2002).
However, there are still 6 optical clusters that do not appear to have
X-ray counterparts. This could be a hint that these clusters are
either the results of projection effects, or that our XMM survey is
too shallow to reveal the probably weak X-ray emission from these
clusters.

In order to address this final issue and to study the relation 
between the limiting flux of our 
X--ray survey with respect to exposure time, we have carried out the following 
experiment. We have analyzed observations taken from 15 XMM public fields with 
mean exposure times $\sim 21$ ksec, 
after filtering to correct for the high particle background  
in the soft 0.3-2 keV band. 
Using the parameters of the SAS software as described previously, we have 
detected 31 candidate clusters. The faintest cluster detected, with a
flux of $\sim 5 \times 10^{-15}$
erg cm$^{-2}$ s$^{-1}$, was found in the deepest field with an exposure
time of 37 ksec. We then reduce the exposure times 
to a new mean value of $\sim 5$ ksec, similar to our shallow survey,
and find only 9 out of the 31 previously identified candidate 
clusters (29\%), having a limiting flux of $\sim 2.3 \times 10^{-14}$ erg cm$^{-2}$ 
s$^{-1}$. Therefore, had we had deeper XMM observations (by
an average factor of $\sim 5$ in exposure time) we would 
have detected $\sim 13$ X-ray candidate
clusters in the region covered by our shallow XMM survey, which is 
consistent (within $1\sigma$) with the number of our optical cluster
candidates. 

\section{Conclusions}
We have made a direct comparison between optical and X-ray based techniques
used to identify clusters. 
We have searched for extended emission in our shallow XMM-{\it Newton} 
Survey, which covers a $\sim 1.6 \;\;{\rm deg^{2}}$ area (8 out of 9
original XMM pointings) 
near the North Galactic Pole region and we have 
detected 4 candidate X-ray clusters.
Out of the 4 X-ray candidate clusters 3 are common with our optical cluster
list .This relatively, small number of optical  
cluster candidates observed in X-rays
suggest that some of the optical cluster 
candidates are either projection effects or
poor clusters and hence they are fainter in X-rays 
than the limit of our shallow survey 
$f_{x}(0.3-2 keV) \rm \simeq 2 \times 10^{-14}erg ~cm^{-2}~s^{-1}$.
This latter explanation seems to be supported from an analysis of 
public XMM fields with larger exposure times.


\begin{chapthebibliography}{1}

\bibitem[Abell, Corwin \& Olowin 1989]{aco89}
Abell, G.O., Corwin, H.G., Olowin, R.P., 1989, ApJS, 70, 1

\bibitem[Couch et al. 1991]{couch91}
Couch, W.J., Ellis, R. S., MacLaren, I., Malin, D. F., 1991, MNRAS, 249, 606

\bibitem[Dalton et al. 1994]{da94}
Dalton, G.B., Efstathiou, G., Maddox, S. J., 
Sutherland, W. J., 1994, MNRAS, 269, 151

\bibitem[Ebeling et al. 2000]{eb00}
Ebeling, H., et al., 2000, ApJ, 534, 133

\bibitem[Frenk et al. 1990]{fre90}
Frenk, C. S., White, S. D. M., Efstathiou, G., Davis, M., 1990, ApJ, 351, 10


\bibitem[Goto et al. 2002]{goto02}
Goto, T., et al., 2002, AJ, 123, 1807

\bibitem[Olsen et al. 1999]{ol99}
Olsen, L. F., et al., 1999, A\&A, 345, 681

\bibitem[Postman et al. 1996]{pos96}
Postman, M., Lubin, L. M., Gunn, J. E., Oke, J. B., Hoessel, J. G., Schneider, D. P.,Christensen, J. A., 1996, AJ, 111, 615

\bibitem[Scharf et al. 1997]{sch97}
Scharf, C. A., Jones, L. R., Ebeling, H., Perlman, E., Malkan, M., 
Wegner, G., 1997, ApJ, 477, 79

\bibitem[Stocke et al. 1991]{st91}
Stocke, J. T., Morris, S. L., Gioia, I. M., 
Maccacaro, T., Schild, R., Wolter, A., Fleming, T. A., Henry, J. P., 1991, 
ApJS,76,813

\bibitem[Stoughton et al. 2002]{stou02}
Stoughton, C., et al., 2002, AJ, 123, 485

\end{chapthebibliography}
\end{document}